\documentclass{article}
\usepackage{epsfig}
\newcommand{\bfr}{\begin{flushright}}
\newcommand{\efr}{\end{flushright}}
 
\begin{document}
\title{Bogomol'nyi Equations for Vortices in Born--Infeld--Higgs Systems
}
\author{Kiyoshi Shiraishi\\
Department of Physics, Faculty of Science, Ochanomizu University,\\
1-1 Otsuka 2, Bunkyou-ku, Tokyo 112, Japan\\ 
and\\
Satoru Hirenzaki\\
Department of Physics, Tokyo Metropolitan University,\\ Setagaya-ku,
Tokyo 158, Japan}
\date{International Journal of Modern Physics {\bf A6}
 (1991) pp. 2635--2647
}
\maketitle
\begin{abstract}
We study vortex solutions in the Born--Infeld theory coupled with a
complex scalar field. We show that for a specific form of the ``Higgs''
potential the vortex satisfies a set of Bogomol'nyi-type equations.
Another model, with nonlinear interaction between gauge and Higgs
fields, is also considered. We show how it is derived from a
supersymmetric extension of the Born--Infeld theory with a minimally
coupled complex scalar field.
\end{abstract}

\section{Introduction}
The idea that cosmic strings \cite{1} might provide the seeds for
galaxy formation \cite{2} has attracted the attention of
particle physicists as well as astrophysicists in recent years \cite{3}.
The scenario of cosmic-string theory seems very simple and highly
predictive. Cosmic strings also have direct consequences which can be
checked by astronomical observations \cite{4}.

We find that, in many research papers, cosmic strings are
represented by Nielsen--Olesen vortices \cite{5} in Abelian Higgs
systems. In the model of this type, the $U(1)$
gauge field is not the ordinary electromagnetic one, but considered as a
field associated with a spontaneously broken part in a grand unified
model.

In unification schemes at very high energy such as SUGRA-GUTs \cite{6}
and superstring(-inspired) models \cite{7}, corrections to the field
equations exist in general. They have a mass scale of order of the Planck
mass; above the Planck energy, nonrenormalizable interactions become
important. The effects of such nonrenormalizable correction terms have
been studied in diverse contexts, especially through the investigation
of physics in very early universe.

We can derive effective actions from string theories by various
methods \cite{9}.  From open-string theory \cite{10}, we can get the
Born--Infeld action \cite{11} for gauge fields. Although the open-string
theory was considered as phenomenologically unfavorable,
four-dimensional ones have been investigated by several authors very
recently \cite{12}.

If the nonlinear gauge  action is utilized in a grand unification scheme,
a topological defect, which can possibly be produced in a phase
transition, might have different structure than in the usual model.
Conversely, we can suppose that topological defects and their
consequences in the very early universe might give some clues to physics
on very-high-energy scale.

We must notice that the shape of Higgs potential could be changed and
nonrenormalizable in the unified models with spontaneously broken
symmetry. If we take the possibility into account, we have a great
variety of models, in which vortexlike solutions exit. In this paper, we
restrict ourselves to considering an interesting class of gauge Higgs
models.

We would like to seek gauge Higgs models, in which the vortex-type
solution can be described as a solution to a set of \textit{first-order}
differential equations. Such equations are known as self-dual or
Bogomol'nyi equations \cite{13} in gauge field theories which possess
topological solutions, such as vortices and monopoles.  Bogomol'nyi-type
equations hold only if couplings in a model take specific values.
In general, if the self-dual equations are satisfied, the mass (or the
energy density) of the topological object is determined analytically,
and it often gives a bound for the mass of the object in a more
generic class of models. Furthermore, a solution which represents
multicentered topological objects can be obtained from the analysis of
Bogomol'nyi-type equations\cite{14}.

We choose Born--Infeld action as a gauge-field action, since it has been
motivated by string theories recently \cite{11} and is a unique action
which describes causal propagation of photons, in spite of its
nonlinearity \cite{15}.
We assume a kinetic term of a complex Higgs scalar field as a minimal
one, i.e., a quadratic in covariant derivatives. Thus the choice
of models is attributed to determination of the form of the Higgs
potential. The present work was partly prompted by a recent paper by
Jackiw and Weinberg \cite{16}, in which they obtain a self-dual relation
in a Chern--Simons Higgs system in three-dimensional space--time.

In Sec.~2, we determine the potential and show the vortex solutions
associated with it. In Sec.~3, we examine another possibility of
existence of nonminimal coupling between gauge and Higgs fields. A
model is shown to possess Bogomol'nyi equations which are the same as
those of the usual Abelian Higgs model. The relation to supersymmetry is
studied. Section 4 is devoted to the conclusion.

\section{Bogomol'nyi Equations, Potentials and Vortices}
We start from a coupled system of a complex scalar field and $U(1)$
gauge field which is governed by nonlinear Born--Infeld action
\cite{11} in two (spatial) dimensions. The total action is
\begin{equation}
S=\int d^2\hat{x}\left[\frac{\hat{\beta}^2}{e^2}
\left\{\sqrt{\det\left(\delta_{ij}+\frac{1}{\hat{\beta}}\hat{F}_{ij}
\right)}-1\right\}+(\hat{D}_i\hat{\phi})(\hat{D}^i\hat{\phi})^*+
\hat{V}(|\hat{\phi}|)\right]\,,
\end{equation}
where $\hat{\phi}$ complex scalar field, $\hat{A}_i$ is a $U(1)$ gauge
field, and $\hat{F}_{ij}=\partial_i \hat{A}_j-\partial_j\hat{A}_i$
($i, j=1, 2$) is the field strength. The covariant derivative is defined
as $\hat{D}_i=\partial_i+i\hat{A}_i$.

We require that the Higgs potential $\hat{V}(|\hat{\phi}|)$ have a
minimum at $|\hat{\phi}|=v$. Further, we assume normalization so that
$\hat{V}(v)=0$.

Rescaling the dimensions of the fields and length by using $v$ 
according to
\begin{eqnarray}
\hat{\phi}&=&v\phi\,,
\label{(2-2a)}\\
\hat{A}_i&=&evA_i\,,
\label{(2-2b)}\\
\hat{x}^i&=&(ev)^{-1}x^i
\label{(2-2c)}
\end{eqnarray}
leads to
\begin{equation}
S=v^2\int d^2{x}\left[{\beta}^2
\left\{\sqrt{\det\left(\delta_{ij}+\frac{1}{{\beta}}{F}_{ij}
\right)}-1\right\}+({D}_i{\phi})({D}^i{\phi})^*+
{V}(|{\phi}|)\right]\,,
\end{equation}
where $\beta=\hat{\beta}/(e^2v^2)$. The rescaled potential $V(|\phi|)=
(e^2v^4)^{-1}\hat{V}(|\hat{\phi}|)$ has a minimum at $|\phi|=1$.

When the $U(1)$ symmetry is broken in the above system, we expect
the existence of classical vortex solutions with finite energy (in two
dimensions). In the case of the usual Maxwell action, a critical relation
between the gauge coupling constant and the Higgs self-coupling leads to
the vortex solution's obeying Bogomol'nyi-type equations \cite{13,17}.
In this section, we show that the vortex in the Born--Infeld Higgs system
satisfies Bogomol'nyi-type equations for a specific choice or the Higgs
potential.

Here we take the simplest Ans\"atze:
\begin{equation}
\phi=f(r) e^{in\theta}\,,\qquad
A_i=-\varepsilon_{ij}\frac{x^j}{r^2}n[P(r)-1]\,,
\label{2.4}
\end{equation}
where $r$ is the radial distance form the origin (i.e., the center of the
vortex), $\theta$ is the azimuthal angle, $\varepsilon_{ij}$ is the
Levi-Civita tensor, and $n$ corresponds to the winding number of the
configuration. We consider $n$ as a positive integer throughout this
paper, since the case with negative winding number can be immediately
constructed from the positive-$n$ case. If the energy of the vortex is to
be finite, $|\phi|=f$ must tend to one and $D_i\phi$ must vanish at
spatial infinity. Of course, the fields must not be singular also at the
origin. Therefore the boundary conditions for a vortex solution are 
\begin{eqnarray}
& &P(0)=1\,,\qquad f(0)=0\,,
\label{(2-5a)}\\
& &P(\infty)=0\,,\qquad f(\infty)=1\,.
\label{(2-5b)}
\end{eqnarray}
Hereafter we study the field equations in terms of these cylindrical
Ans\"atze and boundary condition.

The field equations for the scalar and gauge fields become
\begin{eqnarray}
& &\frac{(rf)'}{r}=\frac{n^2P^2f}{r^2}+\frac{1}{2}\frac{\partial
V(f)}{\partial f}\,,
\label{2.6a}\\
&
&\left[\left\{1+\left(\frac{nP'}{\beta
r}\right)^2\right\}^{-1/2}\frac{P'}{r}\right]'=\frac{2f^2P}{r}\,,
\label{2.6b}
\end{eqnarray}
when the prime denotes the derivative with respect to $r$.

Our purpose is to find differential equations of first order whose
solutions automatically satisfy the above field equations, and to find a
suitable form of the symmetry breaking potential. We begin with the
relation concerned with the covariant derivative on the scalar field.

Since we assume that the kinetic term of the complex scalar is the
same as that of the usual Abelian Higgs model, we may anticipate that one
of the first order self-dual equations is
\begin{equation}
D_1\phi+iD_2\phi=0\,,
\label{2.7}
\end{equation}
just as in the case of vortices in Abelian Higgs model \cite{14}.
Substitution of the Ans\"atze into Eq.~(\ref{2.7}) gives
\begin{equation}
f'=\frac{nPf}{r}\,.
\label{2.8}
\end{equation}

Using this equation twice in the field equation (\ref{2.6a}), we can
obtain a first-order differential equation,
\begin{equation}
\frac{nP'}{r}=\frac{1}{2f}\frac{\partial V(f)}{\partial f}\,.
\label{2.9}
\end{equation}

\begin{figure}[ht]
\begin{center}
\includegraphics[width=5cm]{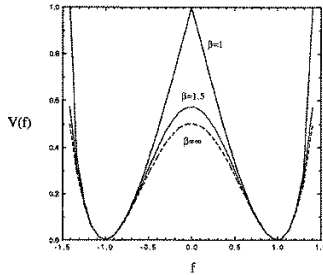}
\caption{Schematic view of the potential $V(\phi)$. The solid, dotted and
dashed lines correspond to $\beta=1, 1.5$, and $\infty$, respectively.
}
\label{f1}
\end{center}
\end{figure}

We determine the potential by demanding that the first order
equations (\ref{2.9}) and (\ref{2.8}) satisfy another field equation,
(\ref{2.6b}). Performing integration after substituting the first two
equations into the third, we obtain
\begin{equation}
\frac{1}{2f}\frac{\partial V(f)}{\partial f}=
\frac{f^2-1}{\sqrt{1-\frac{(f^2-1)^2}{\beta^2}}}\,,
\label{2.10}
\end{equation}
since we require that the vanishing minimum of the potential be
located at $f=1$. Once more, integration leads to the result, i.e.
\begin{equation}
V(f)=\beta\left\{1-\sqrt{1-\frac{(f^2-1)^2}{\beta^2}}\right\}\,,
\label{2.11}
\end{equation}
where the minimum value of the potential is normalized to be zero. The
schematic view of the potential is displayed in Fig.~\ref{f1}. Evidently,
we can replace $f$ by $|\phi|$ to get the potential of complex scalar
field. It is obvious that this potential reduces to the usual
$\phi^4$-type potential in the limit of $\beta\rightarrow\infty$.

The potential cannot be defined in the region
\begin{equation}
(|\phi|^2-1)^2\ge\beta^2\,.
\label{2.12}
\end{equation}
For our present purpose to obtain vortex solutions, it is sufficient
that the potential is defined as the form of Eq.~(\ref{2.11}) in the
range $0\le f\le 1+\epsilon$, where $\epsilon$ is an arbitrary small
quantity. Hence, we restrict $\beta$ to be greater than one, and
$V(\phi)$ can be regarded as an arbitrary function of $|\phi|$ for
$|\phi|>1$.

It can be shown that the Higgs- and the gauge-field masses are equal,
i.e.
\begin{eqnarray}
m_H^2&=&\frac{1}{2}\left|\frac{\partial~2V}{\partial
f^2}\right|_{f=1}\cdot(ev)^2=2e^2v^2\,,
\label{2.13a}\\
m_A^2&=&\left|2f^2\right|_{f=1}\cdot(ev)^2=2e^2v^2\,.
\label{2.13b}
\end{eqnarray}

Now we look into the behavior of vortex solutions. At large $r$,
Eq.~(\ref{2.8}) and (\ref{2.9}) with (\ref{2.10}) can be approximated by
linear equations in terms of $P$ and $\delta\equiv 1-f$. The
linearized equations are the same as those of the Abelian Higgs model
and are therefore independent of $\beta$. The behavior of solution at
infinity is
\begin{eqnarray}
f&\simeq&1-C K_0(\sqrt{2}r)\,,
\label{2.14a}\\
P&\simeq&C\frac{\sqrt{2}r}{n}K_1(\sqrt{2}r)\,,
\label{2.14b}
\end{eqnarray}
where $K_0$ and $K_1$ are the modified Bessel functions.

By performing a power-series expansion, we obtain the behavior at
small $r$. To the leading order, the solution is (when $\beta\ne 1$)
\begin{eqnarray}
f&=&\frac{A(\sqrt{2}r)^{n+2}}{4(n+2)\sqrt{1-\frac{1}{\beta^2}}}+
O[(\sqrt{2}r)^{n+4}]\,,
\label{2.15a}\\
P&=&1-\frac{(\sqrt{2}r)^{2}}{4n\sqrt{1-\frac{1}{\beta^2}}}+
O[(\sqrt{2}r)^{4}]\,.
\label{2.15b}
\end{eqnarray}

\begin{figure}[ht]
\begin{center}
\includegraphics[width=5cm]{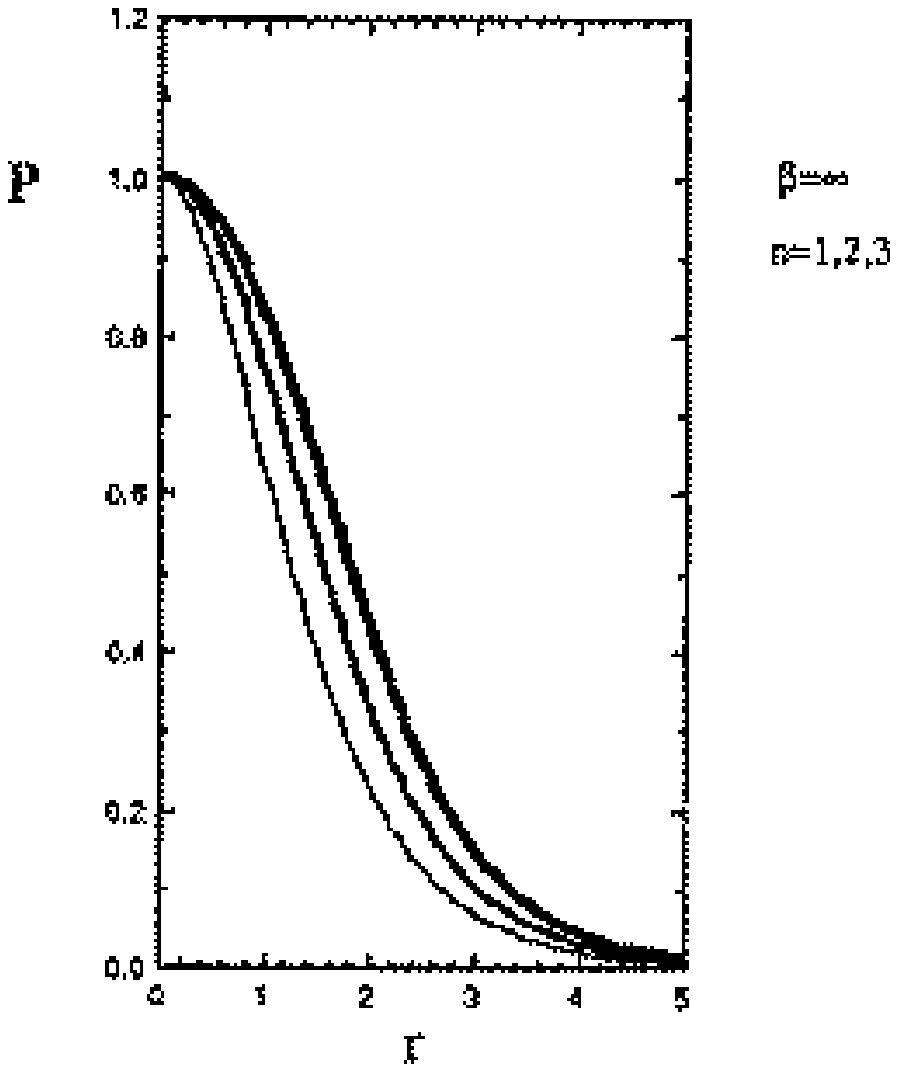}
\includegraphics[width=5cm]{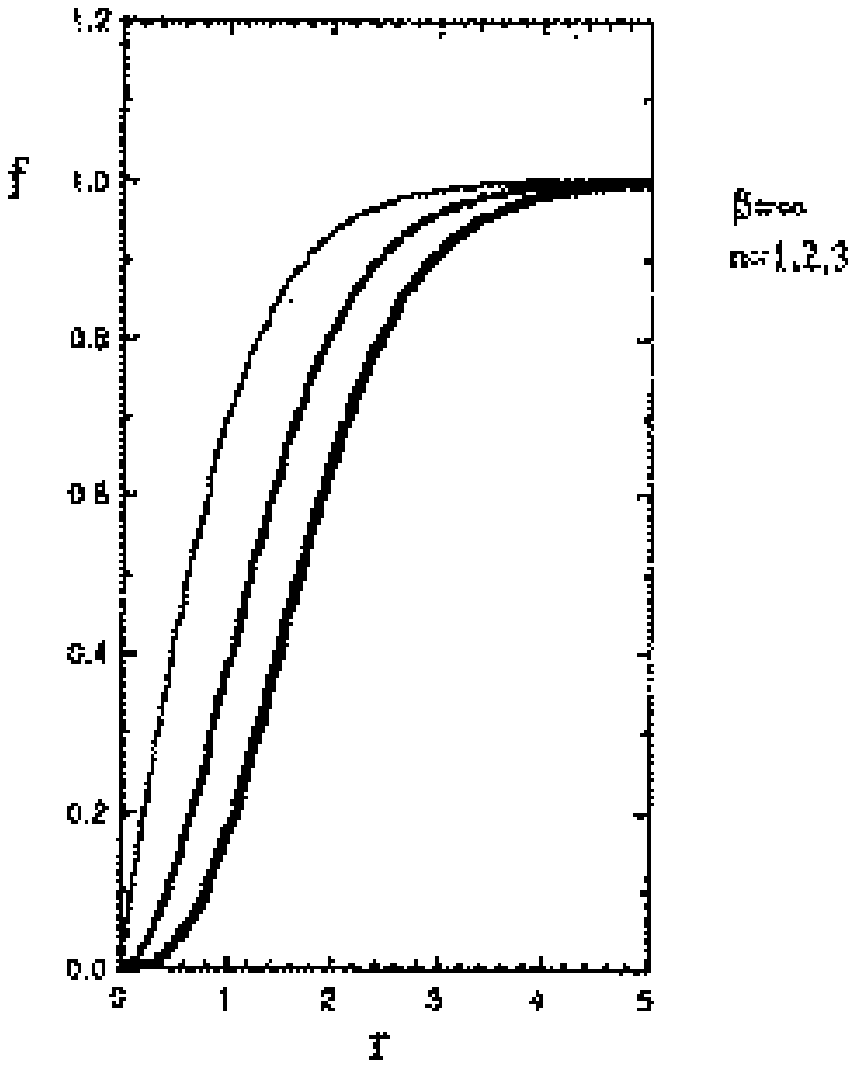}\\
\centering{(a)\hspace{5cm}(b)}
\caption{The functions $P(r)$ [in (a)] and $f(r)$ [in (b)] in the case of
$\beta=\infty$.
The lines correspond to the winding numbers $n=3, 2$ and $1$, in
order of thickness.
}
\label{f2}
\end{center}
\end{figure}

\begin{figure}[ht]
\begin{center}
\includegraphics[width=5cm]{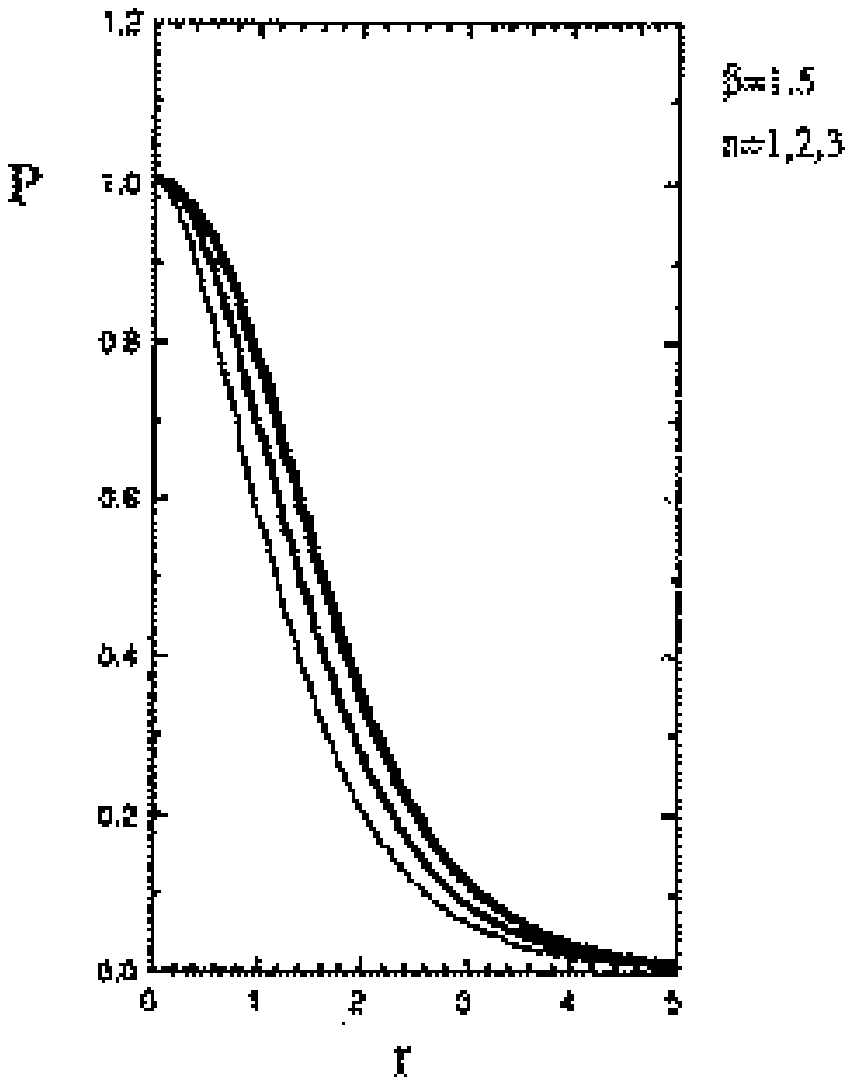}
\includegraphics[width=5cm]{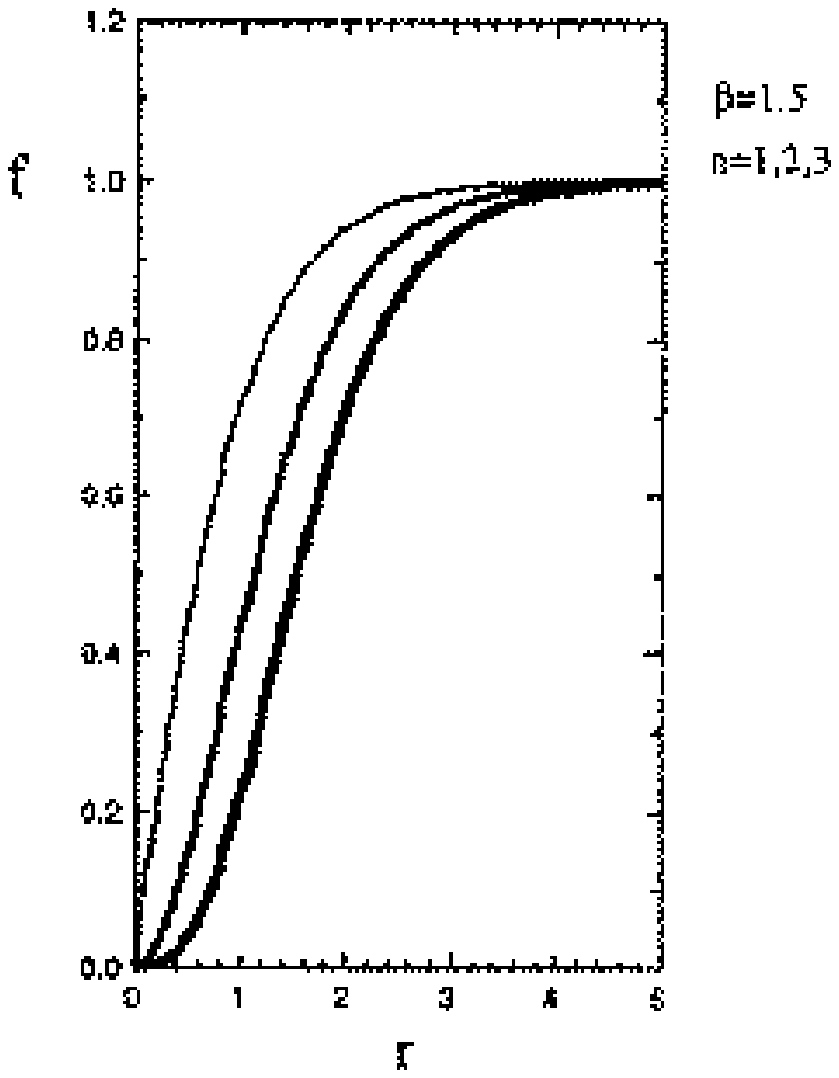}\\
\centering{(a)\hspace{5cm}(b)}
\caption{The functions $P(r)$ [in (a)] and $f(r)$ [in (b)] in the case of
$\beta=1.5$.
The lines correspond to the winding numbers $n=3, 2$ and $1$, in
order of thickness.
}
\label{f3}
\end{center}
\end{figure}

\begin{figure}[ht]
\begin{center}
\includegraphics[width=5cm]{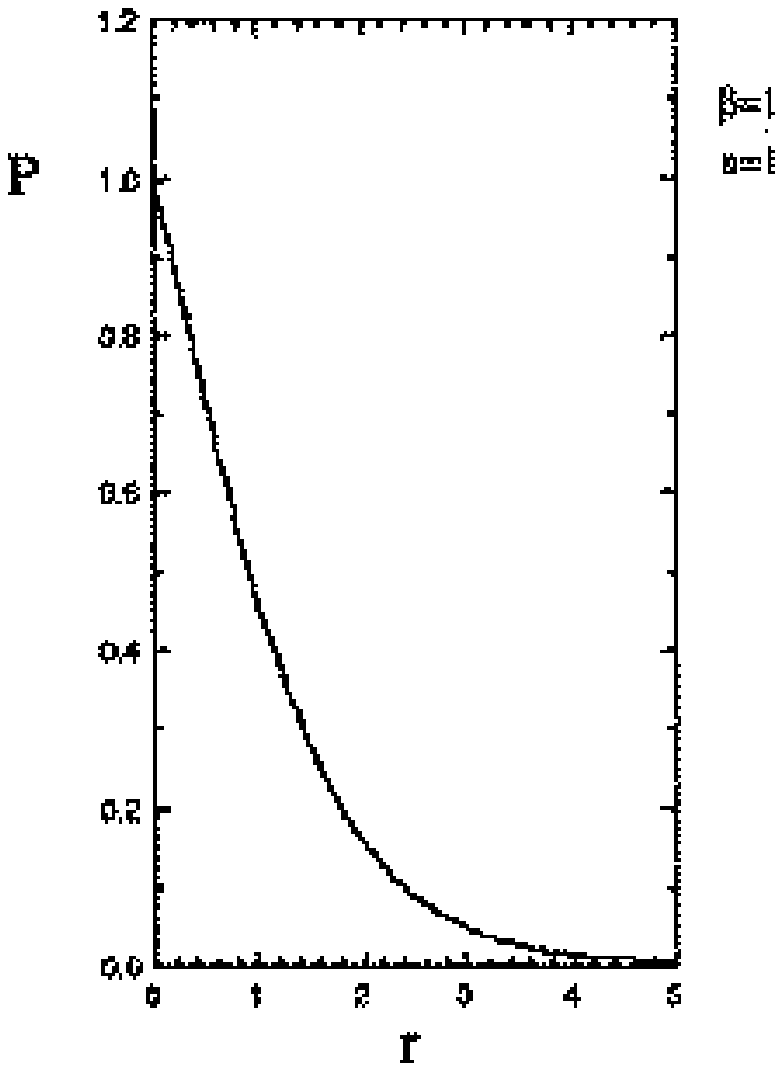}
\includegraphics[width=5cm]{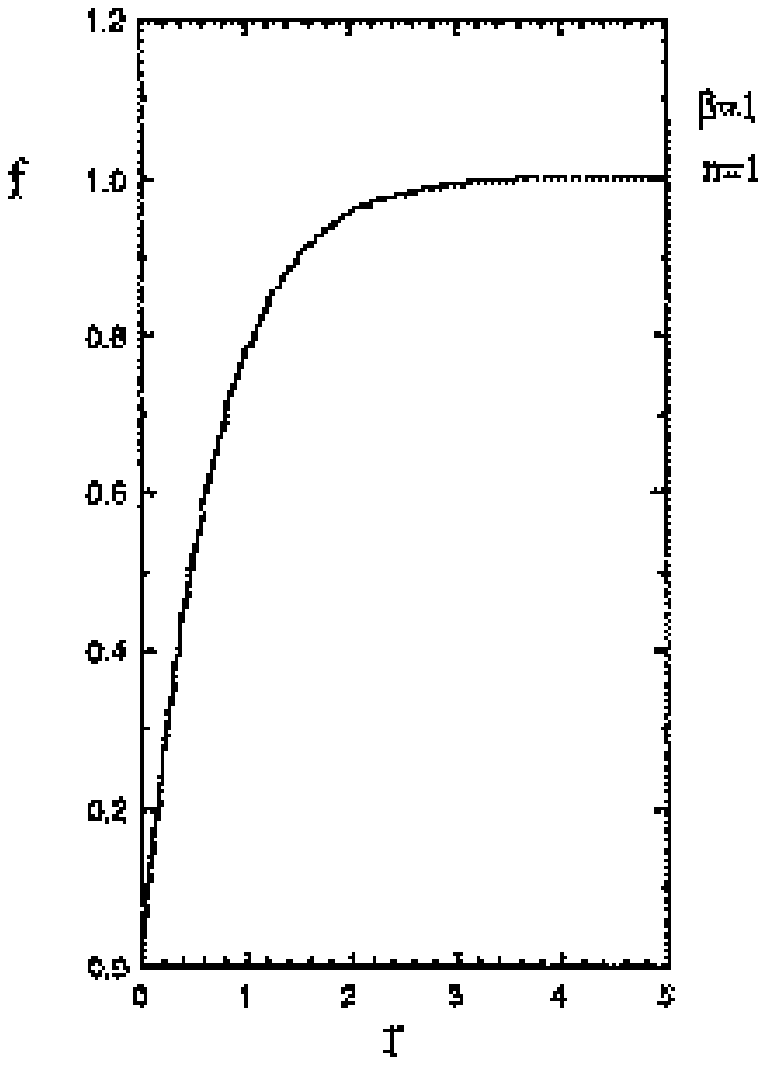}\\
\centering{(a)\hspace{5cm}(b)}
\caption{The functions $P(r)$ [in (a)] and $f(r)$ [in (b)] in the case of
$\beta=1$.
The lines correspond to the winding number $n=1$.
}
\label{f4}
\end{center}
\end{figure}

The constants $C$ [in (\ref{2.14a}) and (\ref{2.14b})] and $A$ [in
(\ref{2.15a}) and (\ref{2.15b})] are not determined by the expansions in
their proper regions. The detailed behavior of vortex configurations is
obtained by numerical calculation of the set of Bogomol'nyi-type
equations. We use the computational code COLSYS \cite{18}. The
results of the numerical solution for the equations are shown in
Figs.~\ref{f2}, \ref{f3} and \ref{f4}, for $\beta=\infty, 1.5$ and $1$
respectively.

The limiting case with $\beta=1$ attracts much interest. As we have
seen, the behavior of the solutions very distant from the core of
the vortex is independent of $\beta$. Hence we pay attention to 
examination of the behavior in the vicinity of the origin. Absence of the
singularity in the vector field at the origin requires $P(0)=1$. Thus $f$
behaves like $r^n$ near the origin. The $P-1$ must behave like
$r^{2-n}$, according to Eq.~(\ref{2.8})--(\ref{2.10}). $P$ can converge
to $1$ if and only if $n=1$, and then we have a finite magnetic flux.
Unlike the case of general values of $\beta$, the merger of two-vortices
into a vortex with $n=2$ is impossible. The results of the numerical
solutions of the equations are shown in Fig.~\ref{f4} for $\beta=1$ and
$n=1$.

The energy of the vortex is obtained by the substitution of the solution
into the integration:
\begin{equation}
E=v^2\int d^2x\left[\beta^2\left\{\sqrt{1+\frac{n^2{P'}^2}{\beta^2r^2}}
-\sqrt{1-\frac{(f^2-1)^2}{\beta^2}}
\right\}+
{f'}^2+\frac{n^2P^2f^2}{r^2}\right]\,.
\label{2.16}
\end{equation}
We rearrange the expression (\ref{2.16}) to give
\begin{eqnarray}
& &E=v^2\int d^2x\left[\left(f'-\frac{nPf}{r}\right)^2+\frac{1}{2}
\frac{1}{\sqrt{1+\left(\frac{n{P'}}{\beta r}\right)^2}}
\left\{\frac{nP'}{r}-\sqrt{1+\left(\frac{n{P'}}{\beta r}\right)^2}
(f^2-1)\right\}^2\right.\nonumber \\
& &+\left.\frac{1}{2}
\frac{\beta^2}{\sqrt{1+\left(\frac{n{P'}}{\beta r}\right)^2}}
\left\{\sqrt{1+\left(\frac{n{P'}}{\beta r}\right)^2}
\sqrt{1-\frac{(f^2-1)^2}{\beta^2}}-1\right\}^2+
\frac{1}{r}\{nP(f^2-1)\}'\right]\,,
\end{eqnarray}
with
\begin{equation}
E\ge 2\pi n v^2\,.
\label{2.18}
\end{equation}
The bound is attained if the self-dual equations, (\ref{2.8}) and
(\ref{2.9}), are satisfied.

If the vortex is embedded in three-dimensional space, it represents an
infinite, straight cosmic string. In this case, the tension of the string
is given by $E$ obtained above. Of course, $E$ is also the energy density
per unit length of the cosmic sting. Moreover, we find that, owing to the
Bogomol'nyi-type equations, all the other components of the stress
tensor, i.e. pressures in the direction perpendicular to the string,
vanish.

\section{Nonlinear Coupling between Gauge and Higgs
Field and Supersymmetry}
In this section, we construct another model in which self-dual vortices
exist. In the present analysis, we permit nonminimal coupling between
gauge and Higgs fields while the kinetic term of the complex scalar is
still assumed to be the canonical one. To be more concrete, we assume a
multiplicative form of modified Born--Infeld action,
\begin{equation}
S=\int d^2\hat{x}\left[\frac{\hat{\beta}^2}{e^2}
\left\{\hat{G}(|\hat{\phi}|)\sqrt{-\det\left(\eta_{mn}+
\frac{1}{\hat{\beta}}\hat{F}_{mn}
\right)}-1\right\}+(\hat{D}_m\hat{\phi})(\hat{D}^m\hat{\phi})^*\right]\,,
\label{3.1}
\end{equation}
where $\hat{G}(|\hat{\phi}|)$ is a function of the Higgs scalar
field $\hat{\phi}$ and $m, n$ are the suffixes of $d$-dimensional
space--time.

Here we require that $\hat{G}(|\hat{\phi}|)$ have an extremum at
$|\hat{\phi}|= v$, and after symmetry breaking the action (\ref{3.1})
becomes the standard Born--Infeld action.

The analysis of the condition on $\hat{G}$ for existence of self-dual
equations can be performed in a way similar to that in Sec.~2. 
We show only the result here. The self-dual equations can be established
if we choose $\hat{G}(|\hat{\phi}|)$ as
\begin{equation}
\hat{G}(|\hat{\phi}|)=\sqrt{1+
\frac{e^4(|\hat{\phi}|^2-v^2)^2}{\hat{\beta}^2}}\,.
\label{3.2}
\end{equation}
Note that in the limit $\beta\rightarrow\infty$, the action reduces to
the usual Abelian Higgs model. The Higgs- and gauge-field masses are
the same as those in the model of Sec.~2.

When we rescale the fields and length as in Sec.~2, and use the same
Ans\"atze, (\ref{2.4}) for a vortex, Bogomol'nyi-type equations are
expressed as
\begin{eqnarray}
f'&=&\frac{nPf}{r}\,,\\
\frac{nP'}{r}&=&f^2-1\,.
\end{eqnarray}

These equations are perfectly coincident with those of the usual
Abelian Higgs model with critical relation between couplings.

The energy in two-dimensional subsystem can be written as
\begin{eqnarray}
& &E=v^2\int
d^2x\left[\beta^2\left\{\sqrt{1+\frac{(f^2-1)^2}{\beta^2}}
\sqrt{1+\frac{n^2{P'}^2}{\beta^2r^2}}
-1\right\}+
{f'}^2+\frac{n^2P^2f^2}{r^2}\right]
\nonumber \\
& &=v^2\int d^2x\left[\left(f'-\frac{nPf}{r}\right)^2+\frac{1}{2}
\frac{\sqrt{1+\frac{(f^2-1)^2}{\beta^2}}}
{\sqrt{1+\left(\frac{n{P'}}{\beta r}\right)^2}}
\left\{\frac{nP'}{r}-\frac{\sqrt{1+\left(\frac{n{P'}}{\beta
r}\right)^2}}{\sqrt{1+\frac{(f^2-1)^2}{\beta^2}}}
(f^2-1)\right\}^2\right.\nonumber \\ 
& &+\left.\frac{1}{2}
\beta^2\frac{\sqrt{1+\frac{(f^2-1)^2}{\beta^2}}}{
\sqrt{1+\left(\frac{n{P'}}{\beta r}\right)^2}}
\left\{\frac{\sqrt{1+\left(\frac{n{P'}}{\beta r}\right)^2}}
{\sqrt{1+\frac{(f^2-1)^2}{\beta^2}}}-1\right\}^2+
\frac{1}{r}\{nP(f^2-1)\}'\right]\,.
\end{eqnarray}
which is the same as that of the model in Sec.~2; and this is an
identical relation to the bound in the usual Abelian Higgs model.

Note that, in the present model, even energy-density distribution of the
vortex solution is independent of $\beta$. The pressure in the direction
perpendicular to the straight string vanishes as in the previous model.

From now on, we will show how the special form of the action of the
present model arises very naturally from a supersymmetric
generalization of Born--Infeld theory. We discuss here an
indirect evidence for connection between our model and supersymmetry.
We will show the complete argument in future publications.

We start with construction of $d=4$, $N=1$ supersymmetric
generalization of the Born--Infeld theory. A general discussion can be
found in Ref.~\cite{19}, on the case without matter couplings. In the
present investigation, we consider a coupling to a complex scalar field,
but we begin with a simplified action. If only magnetic fields are
present in two-dimensional subsystem, say, the $x$-$y$ plane, then the
Born--Infeld Higgs action is equivalent to
\begin{equation}
\int d^4\hat{x}\left[\frac{\hat{\beta}^2}{e^2}
\left\{\sqrt{1+\frac{1}{2\hat{\beta}^2}\hat{F}_{mn}\hat{F}^{mn}}-1\right\}
+(\hat{D}_m\hat{\phi})(\hat{D}^m\hat{\phi})^*\right]\,,
\label{3.6}
\end{equation}
where $m$ and $n$ run over $0, 1, 2, 3$. In present
four-dimensional case, this replacement of action is equivalent to
omitting $(F^*F)^2$ term \cite{19}, which has no effect on dynamics of
the two-dimensional subsystem.

Let us consider a supersymmetric generalization of the action (\ref{3.6})
and its coupling to a complex scalar field with symmetry breaking. To
this end, it is most convenient to use the superfield formalism. The
curvature supermultiplet $W_\alpha$, whose  components are $A_m,
\psi, D$ ($\psi$ is a photino field and $D$ an auxiliary one), is
given by
\begin{equation}
W_\alpha=-\frac{1}{4}\bar{D}^\beta\bar{D}_\beta D_\alpha V\,,
\label{3.7}
\end{equation}
where $V$ is a vector multiplet. We can also construct
$W_{\alpha\beta}\equiv D_\alpha W_\beta$, though $\bar{D}_\alpha
W_\beta=0$. For the expression for the covariant derivative, see
Ref.~\cite{20}. The multiplet $W^\alpha D_\alpha
V+\bar{W}^\alpha\bar{D}_\alpha V$, built from the superfields, contains
the term $2D^2-F^2$ in its last ($D$) component. Here $D$ is an auxiliary
field. On the other hand, we find that the superfield $W\alpha
W_\alpha+\bar{W}^\alpha\bar{W}_\alpha$ has the combination $2D^2-F^2$ as
the first component. Therefore we can construct any supersymmetric
gauge model whose action is an arbitrary function of $F^2$, by
expanding the function in a form of power series and using these
superfields.

To introduce a complex scalar field, we use two chiral scalar
multiplets, $\Phi$ and $\Phi^+$. The scalar kinetic term is
involved in the combination $\Phi^+e^{2V}\phi$ as its last component.
(Note that the gauge coupling is absorbed in the fields in our notation.)
It also includes the coupling between the auxiliary field $D$ and the
scalar field.

Further we introduce a Fayet--Illiopoulos term \cite{21} to break the
$U(1)$ symmetry. This term is linear in the superfield $V$.

Putting the above terms together, we obtain a supersymmetric
generalization of the Born--Infeld theory with symmetry breaking. If we
set the fermion fields to zero, the action for the bosonic fields is:
\begin{equation}
S\approx\int d^4{x}\left[{\beta}^2
\left\{\sqrt{1+\frac{1}{{\beta}^2}\left(\frac{1}{2}F^2-D^2\right)}
-1\right\}
+({D}_m{\phi})({D}^m{\phi})^*+(|\phi|^2-1)D\right]\,,
\label{3.8}
\end{equation}
where the dimensionful quantities have been properly scaled again.

Note here that the gauge action is obtained by the replacement
$^2\rightarrow F^2-2D^2$ in the Born--Infeld action. Eliminating the
auxiliary field $D$ by using its field equation, we get
\begin{equation}
S\approx\int d^4{x}\left[
{\beta}^2
\left\{\sqrt{1+\frac{(|\phi|^2-1)^2}{{\beta}^2}}
\sqrt{1+\frac{1}{{\beta}^2}\frac{1}{2}F^2}
-1\right\}
+({D}_m{\phi})({D}^m{\phi})^*
\right]\,.
\label{3.8}
\end{equation}
This action is precisely equal to (\ref{3.1}) with (\ref{3.2}) up to
normalization.

The intimate connection between self-duality and supersymmetry has
been studied in various models \cite{22}. The both are related to the
existence of fermionic zero modes. We must clarify the coupling of
fermion fields to the action and the full supersymmetric extension of the
Born--Infeld theory which contains general higher-order terms, such
as $(F^*F)^2$ \cite{27}.

\section{Conclusion}
In this paper, we have studied the nonlinear gauge Higgs models 
whose vortex solutions are governed by a set of first-order equations.

In the model shown in Sec.~2, the vortex solutions are slightly
modified in their shape for finite $\beta$. For smaller $\beta$, the
``string'' reduces its thickness, while its energy per unit length is
still unchanged.
If we apply the model to bosonic superconducting strings \cite{23},
the allowed parameter region for existence of the solution will be
enlarged because of the higher energy density of the false vacuum.
Although the calculation of magnitude of critical current or other
characteristic quantities is necessary, no remarkable effect on
properties of possible cosmic strings may be expected because of the
restriction on the range of the coupling, $\beta>1$.

We did not consider the analysis of multivortex configuration. This
is an interesting subject to study, especially in the case of
$\beta\approx 1$.

A generalization to non-Abelian gauge models in very attractive
\cite{24}. Mechanisms of gauge symmetry breaking must be studied in the
Born--Infeld-type theory. At the same time, we wish to study topological
defects in the model, such as monopoles and strings. We are also
interested in the interrelation between the gauge sector and the axionic
sector in the string-inspired models. In particular, we hope to study the
structure of comic strings in such models. Recently, the present authors
and Nakamula have considered gauge-field cosmic strings with
compactified space \cite{25}. Applying such a method to our
Born--Infeld Higgs system, we might clarify the existence of several
types of cosmic strings and their implications for cosmology. 

In this paper we adopted a minimal kinetic term for a complex scalar.
It is not unnatural to consider noncanonical kinetic term like a
nonlinear sigma model. This type of the kinetic term frequently
appears in the literature of supergravity theory \cite{6,26}. The
consideration of this possibility might allow variety of the form of the
potential which leads to self-dual equations.

The nonlinear model treated of in Sec.~3 will be carefully examined
further. Not only global but local supersymmetry must be considered,
since we also wish to study the relation to unified, phenomenologically
viable models.

\section*{Acknowledgments}
The authors would like to thank A.~Nakamula for some useful
comments. K.~S. would like to thank A.~Sugamoto for
reading this manuscript.

K.~S. is indebted to Soryuusi Shogakukai for financial support. He also
would like to acknowledge the financial aid of Iwanami F\=ujukai.


\end{document}